\documentclass[conference]{IEEEtran}

\usepackage{cite}

\ifCLASSINFOpdf
  \usepackage[pdftex]{graphicx}
\else
  \usepackage{graphicx}
\fi
\usepackage{dblfloatfix}

\usepackage{url}
\usepackage{array}
\usepackage{tabularx}
\usepackage{float}

\usepackage[colorlinks=true,linkcolor=blue,citecolor=blue,urlcolor=blue]{hyperref}

\begin{document}
\title{When ``Do Not'' Is Not \texttt{deny}: Security Rules in \texttt{CLAUDE.md} vs.\ Built-In Controls}

\author{\IEEEauthorblockN{Ting Yan}}

\maketitle

\begin{abstract}
In \texttt{CLAUDE.md}, ``do not'' is a natural-language instruction that the model interprets. Claude Code's \texttt{deny} is a built-in control that blocks an action before the agent can take it. Both can express the same security goal, but they control the agent in different ways. We measure this gap in 481 public \texttt{CLAUDE.md} files. An LLM matched the extracted candidate rules against Claude Code's documented controls, and two security practitioners independently checked a sample without seeing the model's answers or each other's labels. Depending on how closely a control had to match the written rule, only about 4--16\% of the retrieved security rules had a matching built-in control. Under the strictest standard the estimate was 4.4\% (95\% CI: 2.6--6.7\%), and the two annotators agreed closely on which rules had a match. A manual review of complete files found that our extraction method captured 66.3\% of eligible security rules; the reported rates therefore apply to the rules it captured. This is a usable security problem: \texttt{CLAUDE.md} is a write-only channel. A developer writes a security rule but gets no feedback on whether a control will enforce it. The same plain-text form hides two kinds of rule: those a permission rule, mode, or sandbox can enforce, and those left to the model to interpret.
\end{abstract}

\IEEEpeerreviewmaketitle

\section{Introduction}
Large language models have lowered the barrier to building software. Many people now build applications and coding agents with little security experience, and they increasingly direct those agents in natural language. Agent instruction files such as \texttt{CLAUDE.md} and \texttt{AGENTS.md} extend this to security: a developer writes a rule like ``never write secrets to logs,'' ``ask before deleting production data,'' or ``only push through the review branch,'' in plain prose, and the agent reads it. But this is a write-only channel. In ordinary software, stating a requirement comes with feedback that closes the loop: code compiles or fails, tests pass or fail, and logs show what ran, so a developer can tell whether the requirement took effect. A natural-language security rule returns nothing back, and neither the file nor the agent application signals whether anything will enforce it.

Behind that silence, the rules divide into two kinds. Some match a built-in control, such as a permission rule or a sandbox restriction, that the platform applies directly. Others are only text for the model to interpret, unless a team adds a hook, script, or other check. Claude Code documents the split: \texttt{CLAUDE.md} shapes what the model attempts, whereas permission rules and related controls determine what the platform permits~\cite{anthropic2026permissions}. Prior evidence shows why it matters: shipped command denylists are 69.0 to 98.6\% fragile, and removing one declared-scope sentence raises an agent's out-of-scope action rate from 0.0 to 17.1\%~\cite{rashidi2026}. Rules with and without a matching control share the same natural-language form.

Several groups of studies lead to our question, but none answers it. First, studies of agent instruction files describe what developers write. Agent READMEs~\cite{chatlatanagulchai2025} reports security content in 14.8\% of files, and a study of 12,110 Cursorrules files~\cite{sun2026} also finds security content uncommon. These studies show that security rules appear in real files, but they do not check whether each rule has a matching control. Second, systems such as ContextCov~\cite{sharma2026} and Prose2Policy~\cite{gupta2026} turn natural-language instructions or policies into executable checks or code. They show that a new control can be built from text. Our question comes before that step: does Claude Code already provide a built-in control for the rule? Third, studies of agent permission systems examine which controls products provide and how users apply them~\cite{michael2026}. They start from controls that already exist, rather than from security rules that developers have written. As a result, they do not tell us how many written rules can use an existing control and how many remain text for the model. A recent review also identifies the writing of agent security rules as an open problem~\cite{rashidi2026}. We address one concrete part of it by measuring the fraction of real \texttt{CLAUDE.md} security rules that match a built-in Claude Code control without extra code.

We collected 481 public repositories that contain \texttt{CLAUDE.md}. We split each file into lines or sentences and kept segments with predefined rule words such as ``must,'' ``never,'' or ``do not.'' An LLM answered two questions for each segment. Could ignoring it cause a security or privacy harm? If so, does Claude Code already provide a built-in control that can apply the same rule without extra code? Two security practitioners independently reviewed the same 180 segments without seeing the model's answers. We selected 60 segments from each of three groups created by the model: not security-related, security-related without a matching control, and security-related with a matching control. These groups occur at different rates in the full dataset, so we gave each group weight based on its actual size. We used these weights to estimate the share of security rules found by our extraction method that have a matching built-in control.

Consider two rules a developer might write. ``Never run \texttt{npm publish}'' can be matched by a \texttt{deny} permission rule and checked before the command runs; ``never put customer secrets in logs'' needs the model to decide what counts as a secret and whether an output is a log. The first is applied deterministically, the second depends on the model's interpretation, yet both are ordinary sentences the developer writes and moves past. To learn which protection a rule actually has, a developer must leave the file and read the permission and sandbox documentation. So a developer can finish writing a security rule without realizing a separate control is still needed, and a secret can reach a log even though the file says ``never.'' Usable security research treats this as a design problem: security tools should help people make safe choices rather than place all the security work on them~\cite{adams1999users,green2016developers}, and developers are users of these tools too~\cite{acar2016developer}.

The closest study, by Michael and Roesner, examines permission interfaces in five commercial agents, including Claude Code~\cite{michael2026}. It asks what controls each product provides and where those controls fall short. We start from the other side. For each security rule that a developer wrote in \texttt{CLAUDE.md}, we ask whether Claude Code already provides a built-in control that matches it. This different starting point produces different evidence. Michael and Roesner show that commercial agents are missing some controls. We analyze 481 public files and count how often a written security rule has no matching control. Our main result is therefore a percentage, rather than a list of missing product features. We also group the rules by the security issue they address, such as authorization, destructive actions, and secrets.

This paper makes two contributions:
\begin{enumerate}
\item We measure this gap in 481 public \texttt{CLAUDE.md} files. Our extraction method found 4,661 candidate segments, and the classifier marked 870 as security-related. Depending on how closely a built-in control had to match the written rule, only about 4--16\% of the retrieved security rules had a match. We also group the rules by the security issue they address, such as authorization, destructive actions, and secrets.
\item We identify a usable security problem: \texttt{CLAUDE.md} is a write-only security channel. The same natural-language form covers rules the platform enforces and rules left to the model to interpret, and neither the file nor the agent shows which is which, so writing a rule gives no feedback on whether it is enforced.
\end{enumerate}

Our replication package includes the list of sampled repositories, the labeling instructions, and the analysis code.

\noindent\textbf{RQ:} Among the security rules found by our extraction method in public \texttt{CLAUDE.md} files, what percentage have a matching built-in Claude Code control that can apply the same rule without extra code?

\section{Related Work}
\subsection{What developers write in coding agent instruction files}
Developers use files such as \texttt{CLAUDE.md}, \texttt{AGENTS.md}, and Cursorrules to tell coding agents how to work. Prior studies have examined what these files contain. A study of 2,303 agent instruction files from 1,925 repositories found security-related content in 14.8\% of the files~\cite{chatlatanagulchai2025}. A separate study of 12,110 Cursorrules files found that about 4.4\% of the coded instructions concerned security~\cite{sun2026}. Other studies describe the types of configuration developers use~\cite{santos2025,galster2026,jiang2025}, how instruction files grow over time~\cite{chakrabarti2026}, and how developers use them to state values and responsibilities~\cite{treude2026}.

Together, these studies show that developers write security rules for coding agents, although security is not the main subject of most instruction files. They classify what developers write, but they do not check whether the coding agent has a built-in control for each rule. Our study begins with the security rules found in these files and asks whether Claude Code can apply the same rule through a permission, sandbox setting, or another built-in control.

\subsection{Why a written rule is not a security control}
A security instruction written in plain language does not itself block a command, prevent a file write, or stop a network request. The model reads the instruction and decides how to act. Studies have found that repository instruction files do not consistently improve task success~\cite{gloaguen2026,khatri2026}, that some types of instructions are often ignored~\cite{ouatiti2026}, and that adding rules can help even when the benefit does not closely track what the rules say~\cite{zhang2026}. Security research also shows that malicious content can override instructions given to an LLM-integrated application~\cite{greshake2023not}, while tested prompt-injection defenses still leave important failures~\cite{liu2024formalizing,debenedetti2024agentdojo}. These findings show why a sentence interpreted by a model does not provide the same certainty as a rule applied directly by the platform.

Several systems address this problem by turning natural-language rules into executable controls. ContextCov generates checks from coding agent instruction files~\cite{sharma2026}. Prose2Policy generates Rego policy from natural-language access rules~\cite{gupta2026}, and AutoCedar combines generated policy with formal verification~\cite{vatsa2026}. Other work places a separate monitor between an agent and the tools it uses~\cite{palumbo2026,li2026,uppala2026}. These systems show that prose can be converted into a stronger control. Our question comes before that conversion: how often does the coding agent already provide a built-in control for the written rule, without requiring the developer to generate or write additional code?

\subsection{Permission systems and the developer's security work}
Research on agent security has proposed controls over tool arguments~\cite{shi2025}, condition-based runtime rules~\cite{wang2025}, task-specific permissions~\cite{sharma2026b}, filesystem access~\cite{sharma2026c}, execution isolation~\cite{wu2024}, data access~\cite{bagdasarian2024}, and permission graphs~\cite{zhang2026b}. Michael and Roesner survey 21 research proposals and five commercial agents, including Claude Code, and compare how their permission systems are written, produced, and applied~\cite{michael2026}. This work explains what different permission mechanisms can do and where current products lack controls. A recent review also identifies the writing of agent security policy as an open problem~\cite{rashidi2026}.

Our study starts from the developer's side of the problem. For each security rule that a developer has already written, we ask whether an existing Claude Code control can apply that same rule. SkillGuard also studies real agent artifacts, but it measures which protected objects are covered across agent skills rather than whether developers' written security rules have matching controls~\cite{pan2026}. Koch proposes a process for deciding whether a governance rule can be observed and enforced, but does not measure how often built-in controls cover rules found in public repositories~\cite{koch2026}.

This gap has a precedent in usable security. When people write access-control policies in natural language, what they intend often exceeds what the system can represent~\cite{inglesant2008expressions}. Developers writing security rules for coding agents hit the same wall, and the instruction file leaves them to resolve it alone. Our measurement shows how often the platform can resolve it for them, by already providing a matching control.

\subsection{What existing evaluations measure}
Most existing evaluations study what happens after an instruction or control has already been chosen. Instruction-following benchmarks test whether a model obeys rules presented through different instruction files~\cite{huang2026,panavas2026}. Runtime studies test whether controls actually stop an agent; for example, Stop Means Stop examines whether framework stop controls still work during concurrent execution~\cite{khan2026}. Other security studies examine whether operators configure available controls correctly~\cite{dietrich2018investigating} or whether secrets still reach public repositories~\cite{meli2019bad}.

Our study asks an earlier question: does a matching built-in control exist for the written rule? This is separate from whether the developer enabled the control, configured it correctly, or whether it works reliably at runtime. It is also separate from whether the model follows the prose. If a matching control exists, the developer may be able to apply the rule through fixed platform behavior. If no matching control exists, the rule remains dependent on model interpretation unless the developer adds another check.

Studies of policy-as-code also measure whether formal policy languages cover requirements from vendors or regulatory standards~\cite{ruohonen2026,fuchs2025}. They do not start from natural-language security rules written for coding agents. None of the studies above reports what percentage of real, developer-written coding agent security rules have a matching built-in platform control. That percentage is the measurement provided by this paper.

\section{Definitions and Method}
This section explains how we collected the files, found possible rules, classified them, and checked the results.

\subsection{Definitions}
\begin{itemize}
\item \textbf{Security rule}: an instruction in a \texttt{CLAUDE.md} file that tells the coding agent what it must, must not, or may do. We count it as security-related when violating it could expose or alter data, interrupt a system, bypass a permission or required approval, or harm privacy. We exclude general advice about style, building, testing, or workflow, unless breaking it would cause one of these harms.
\item \textbf{Retrieved instruction segment}: a Markdown line or sentence found by our extractor because it contains one of our predefined rule words, such as ``must,'' ``never,'' ``only,'' ``do not,'' or ``ask before.'' If one sentence contains several connected clauses, we keep the whole sentence as one segment.
\item \textbf{Built-in control}: a control documented by Claude Code that a developer can configure without writing executable code. Permission rules and sandbox settings are examples. We ask whether such a control exists, not whether a sampled repository has enabled it or whether it always works at runtime.
\item \textbf{Matching built-in control}: a built-in control matches a written rule only when it covers the whole rule: the same action, the same target, the condition that makes the action wrong, and who must approve it. It is not a match if the only way to apply it is to also block many safe actions.
\item \textbf{Primary label}: a yes-or-no label recording whether a matching built-in control exists for the rule.
\item \textbf{Exploratory secondary category}: for rules without a matching built-in control, the classifier records one possible reason: added code could check it, the model would have to judge each case, or the platform cannot see what it would need to apply the rule. These categories were not checked against manual labels, so we use them only for exploratory analysis (Section~\ref{sec:split}).
\end{itemize}

Our main measurement asks one question: for a security rule found by our extractor, does Claude Code document a built-in control that can apply the same rule without added code?

\subsection{Corpus and sampling}
\noindent\textbf{Sampling frame.} Our primary sample is drawn from GitHub code search for the filename \texttt{CLAUDE.md}, retrieved on 2026-08-14. Code search returns at most 1,000 results per query, ranked by relevance, so this frame is a capped, relevance-ranked slice of public files, not a random sample. We limit our claims to this frame and discuss the limitation in Section~\ref{sec:limits}. We freeze the retrieved pool as a published list and sample from it with a fixed random seed.

\noindent\textbf{Inclusion and pinning.} A single pipeline is applied to every repository: keep one primary \texttt{CLAUDE.md} (root, else \texttt{.claude/CLAUDE.md}); require it be non-empty, at least 500 characters, and English-dominant (at least 92\% of alphabetic characters ASCII); and exclude forks, mirrors, and files with no pinnable commit. Every included file is re-read from its host and pinned to a commit hash, so results do not depend on a mutable branch. We release the frozen pool, the manifest, and the seed (Appendix~\ref{app:repro}).

\noindent\textbf{Replication on a second frame.} Because the primary frame is a convenience sample, we repeat the classifier measurement on an independent, published corpus, the Agent READMEs dataset~\cite{chatlatanagulchai2025}, and report whether the rate and classifier-estimated family pattern recur. Frame B repeats the measurement on a different corpus (a sampling-frame replication), not a second human check.

\begin{figure}[t]
\centering
\includegraphics[width=\columnwidth]{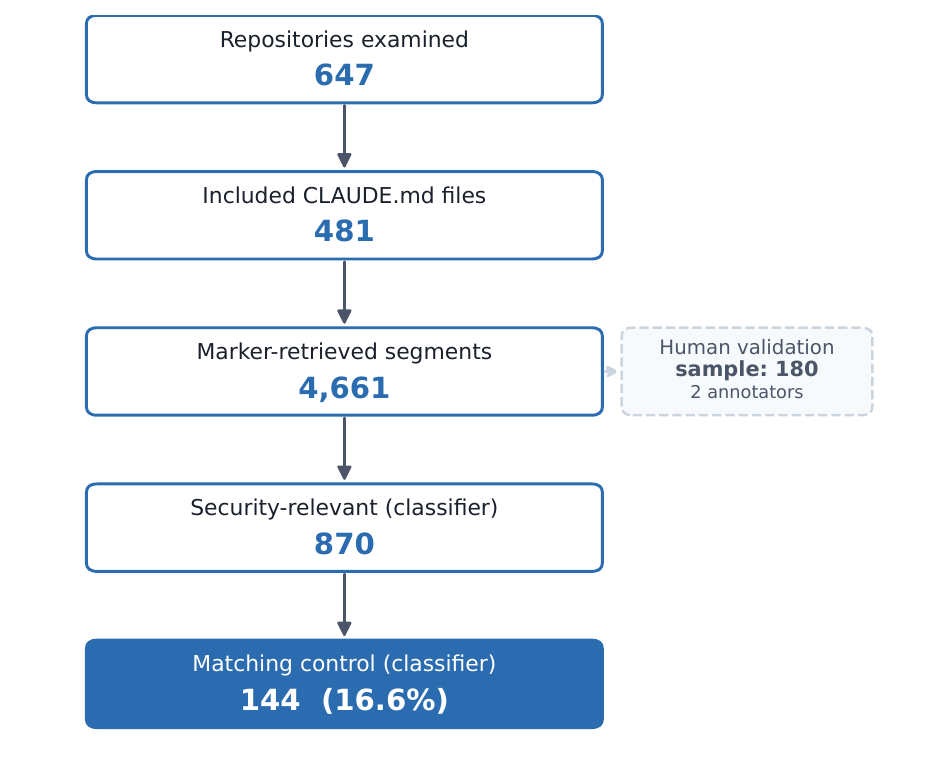}
\caption{Sampling and classification pipeline with per-stage counts.}
\label{fig:pipeline}
\end{figure}

Figure~\ref{fig:pipeline} shows the pipeline with the count at each stage: repositories examined, included files, candidate segments retrieved and classified, security-relevant segments, and those with a matching built-in control, plus the annotator-reference sample.

\subsection{Primary label and secondary categories}
\label{sec:split}
The primary label is binary: does a matching built-in control exist for the rule? One built-in control, or a small set of them, must cover the complete rule. The headline coverage rate depends only on this yes-or-no judgment.

Two real rules show the two labels. ``Do not run \texttt{rails credentials}'' counts as a match: a permission \texttt{deny} rule blocks that exact command. ``Never commit secrets or tokens'' does not: the built-in controls can match commands and paths, but none can read commit content for secrets, so covering it needs an added tool such as a secret scanner or pre-commit hook. Both rules are ordinary prose in the same file, and the file does not record which one has a match.

For rules without a matching built-in control, we record a secondary reason, decided in this order:
\begin{enumerate}
\item \textbf{missing-context}: the control cannot get the information the rule needs, such as where data came from, whether someone approved the action, or the state of another system.
\item \textbf{custom}: the needed information is available and the rule has a clear yes-or-no answer, but applying it takes extra code beyond the built-in controls: a hook, scanner, continuous integration (CI) check, or wrapper, written by hand or generated by a tool such as ContextCov~\cite{sharma2026} or Prose2Policy~\cite{gupta2026}.
\item \textbf{model-mediated}: applying the rule needs an open-ended judgment that cannot be written as a fixed check over the fields the platform can see.
\end{enumerate}
These reasons were not included in the annotator-reference task and therefore have no human-reference validation. We retain them only as exploratory classifier annotations for transparency and hypothesis generation; no percentage or ordering from this split answers the RQ or supports a population claim.

\subsection{Control reference}
Before classification, we froze Claude Code's documented controls at a fixed date and built one short reference table. For each control, the table records what action, target, and condition it can inspect; what it does; its scope; and the dated official source. Built-in controls include \texttt{allow}, \texttt{ask}, and \texttt{deny} rules; tool, command, path, and domain patterns; permission modes; filesystem and network sandboxing; and managed restrictions. Hooks do not count as built-in controls, for two different reasons. A \texttt{PreToolUse} hook only does something if the developer writes executable code or plugs in another decision tool, so it is not a no-code control. Prompt- and agent-based hooks are excluded on the other ground: they decide by model judgment, not by a fixed rule. The classifier and the human coders use the same table, which is based on the official documentation~\cite{anthropic2026permissions,anthropic2026settings,anthropic2026sandbox,anthropic2026hooks} and reproduced in Appendix~\ref{app:control}.

After we finished labeling, we re-checked the frozen table against the official documentation. It was not perfect. We found three small problems, and for each we asked one question: would it have changed any label? First, the entry for \texttt{additionalDirectories} was backwards. It described the setting as restricting which directories a tool may touch, but the setting actually grants access to extra directories. This would matter only if some rule had been counted as covered by \texttt{additionalDirectories}. None was, so no label changed. Second, we had omitted a real built-in feature, \texttt{sandbox.credentials}, which keeps named credential files and environment variables out of reach of sandboxed Bash commands. This would matter only if a credential rule should have counted as covered because of it. We read every security rule that mentioned credentials, plus 39 more rows whose wording came close, and found none within what \texttt{sandbox.credentials} actually governs. So no label changed here either. Third, the network-sandbox entry was imprecise. The sandbox blocks hosts outside the allowlist only when the developer also sets \texttt{strictAllowlist} or a managed lockdown; an \texttt{allowedDomains} list on its own blocks nothing. We therefore count the network sandbox as a match only when the rule calls for that restrictive setting, and any looser crediting is already counted in the error analysis. Because none of the three problems changed a label, we left the frozen table exactly as it was during labeling. We report the audit rather than quietly editing the table after seeing the results, and we release both the table and the audit record with the analysis.

\subsection{Classification and validation}
Extraction supplies candidates, not final labels. A plain parser treats each non-code Markdown line as a segment, splits long paragraph lines at sentence boundaries, and retrieves segments containing predeclared rule words (``must,'' ``never,'' ``only,'' ``do not,'' ``ask before,'' and related forms). Connected clauses within a sentence remain one segment. The reported numbers use only this rule-word pass. Our released code also includes an optional LLM step that can catch more rules, but we did not use it here.

We measure rule-word-retrieval recall with a separate whole-file audit. The annotator, a security practitioner, reads a simple-random sample of 50 included files in full, blind to \texttt{candidates.csv} and classifier outputs, and records every prospective security rule that an agent could violate through its own action. A second scope pass over the recorded rows excludes text that only describes the current architecture or security mechanism. This leaves 95 in-scope reference segments from 135 recorded rows. A reference rule counts as retrieved when the extractor pulled a candidate from the same file that overlaps its lines and closely matches its text. We treat two texts as a close match when they are identical, when one contains the other, or when they share at least 35\% of their words (token Jaccard similarity). The recall interval resamples the 50 audited files as clusters in 10,000 bootstrap replicates.

We then classify every candidate with a large language model (LLM) against the frozen control reference, on two binary judgments: whether it is a security rule, and whether a matching built-in control exists. One frozen prompt (Appendix~\ref{app:prompt}) is applied to all candidates, and the classifier must name the specific built-in control whenever it says a match exists. The prompt tells the classifier to count uncertain cases as matches to discourage false negatives. This does not guarantee the classifier errs in only one direction, so we report the full-corpus classifier estimate separately from the validation-adjusted estimate.

The classifier is a measurement instrument, not the authority. The annotator labels a stratified sample of 180 candidates while blind to the classifier's labels: 60 from each of three classifier strata (not security-related, security-related without a match, and security-related with a match). We call these annotator reference labels, not independent expert labels or ground truth. Because those strata contain 3,791, 726, and 144 population rows, respectively, we use inverse-probability weights for precision, recall, and the validation-adjusted coverage estimate. Confidence intervals resample the 102 sampled repositories as clusters and recompute the post-stratification weights in each of 10,000 replicates. After fixing the binary decisions, the same annotator assigns one of six resource families to the 91 annotator-reference security-positive rows while blind to the classifier's family labels. We report raw and design-weighted exact family agreement, with the weighted interval using the same repository-clustered bootstrap. The annotator's sample contains only 4 egress and 3 personal-data rows, so it is not used to claim stable family-specific population rates. We freeze the prompt before drawing the sample and do not tune it to the annotator labels. The secondary reasons (Section~\ref{sec:split}) are classifier-only outputs and are not included in the annotator-reference task.

A second annotator, also a security practitioner, independently labeled the same 180 candidates, blind to both the classifier's labels and the first annotator's, under the same coding instructions and frozen control reference. We report inter-annotator agreement before adjudication (Section~\ref{subsec:reliability}), then adjudicate every disagreement against the frozen control reference to form the consensus reference used for the classifier's precision and recall and for the coverage estimate. The recall audit above stays single-annotator.

\section{Results}
\subsection{Overall coverage rate}
Most \texttt{CLAUDE.md} security rules have no built-in control that matches them. How many do depends on how strictly the control must match the rule. Table~\ref{tab:coverage} gives the estimate under three standards, from the full-corpus classifier to a strict, adjudicated standard in which a matchable command or path is not enough on its own and the control must actually constrain the agent. Under a looser reading, a control counts as a match if it can merely target the rule's command, path, or domain, even when it does not cover the whole rule; one annotator coded the sample this way, and we include it to show how much the choice of standard moves the number. Because how strictly a control must match, not sampling error, drives the estimate, we give the range, about 4 to 16\%, and treat the strict estimate, 4.4\% [2.6, 6.7], as our primary result: under it, 95.6\% [93.3, 97.4] of security rules have no matching control. These counts come from 4,661 retrieved candidate segments, of which the classifier labels 870 security-related and 144 as having a match. Figure~\ref{fig:pipeline} shows the pipeline counts.

\begin{table}[t]
\centering
\caption{Built-in-control coverage among retrieved security rules, under three matching standards from loose to strict. The classifier rate is over the full corpus (870 segments); the human standards are weighted estimates from the 180-item validation sample. We treat the adjudicated strict standard as our primary estimate. Intervals are repository-clustered 95\% confidence intervals.}
\label{tab:coverage}
\begin{tabular}{lrr}
\hline
Matching standard & Has a match & No match (gap) \\
\hline
Classifier (full corpus) & 16.6\% [12.5, 21.4] & 83.4\% [78.6, 87.5] \\
Human, loose (single annotator) & 14.3\% [7.6, 23.4] & 85.7\% [76.6, 92.4] \\
Adjudicated, strict (primary) & 4.4\% [2.6, 6.7] & 95.6\% [93.3, 97.4] \\
\hline
\end{tabular}
\end{table}

\subsection{Reliability and recall}\label{subsec:reliability}

\noindent\textbf{Reliability.} The full-corpus classifier puts coverage at 16.6\%, well above what the human reference supports. Checked against that reference, the classifier flags every rule the annotators accept as having a control, but also flags many they reject (recall 1.000, precision 0.350; Table~\ref{tab:reliability}); under the strict, adjudicated standard, only 4.4\% have one. The first gate, whether a segment is a security rule at all, is more accurate but still misses about a third of real security rules. Precision and recall are design-weighted against the adjudicated 180 reference labels, 60 from each of the three classifier strata (population weights 63.18, 12.10, and 2.40). The matching standard, not sampling error, mainly moves the estimate: the data do not pin down a single number within the 4 to 16\% range.

\begin{table}[t]
\centering
\caption{Reliability of the two binary judgments (columns). Classifier precision and recall are scored against the adjudicated 180 reference labels, design-weighted. Agreement and Cohen's $\kappa$ compare the two annotators, unweighted: the security-relevant judgment over all 180 items, and the matching-control judgment over the 115 items both annotators judged security-related. All intervals are repository-clustered 95\% confidence intervals.}
\label{tab:reliability}
\footnotesize
\begin{tabular}{lcc}
\hline
 & Security-relevant & Matching control \\
\hline
Classifier precision & 0.894 [0.816, 0.958] & 0.350 [0.219, 0.488] \\
Classifier recall & 0.672 [0.523, 0.867] & 1.000 \\
Annotator agreement & 90.0\% [85.7, 94.0] & 96.5\% [93.0, 99.2] \\
Cohen's $\kappa$ & 0.77 [0.67, 0.86] & 0.89 [0.77, 0.97] \\
\hline
\end{tabular}
\end{table}

\noindent\textbf{Inter-annotator agreement.} Two annotators labeled the 180-item sample independently, each blind to the classifier and to the other. Before adjudication they agree on the matching-control decision for 111 of the 115 rules both judged security-related ($\kappa = 0.89$; Table~\ref{tab:reliability}), and prevalence-robust statistics agree (Gwet's AC1 0.95, positive-specific agreement 0.91), so this is not an artifact of rare positives. They agree less on the prior gate, whether a segment is a security rule ($\kappa = 0.77$), and almost in one direction, with the second annotator more inclusive. We adjudicate all 22 disagreements against the frozen control reference to form the consensus reference used for the estimates; the four contested match rows resolve to three without a control and one with. Because both coders followed the same written guide, we read their high pre-adjudication agreement as showing the labels can be reproduced under that guide, not as proof that the guide itself draws the boundary correctly.

\noindent\textbf{Extraction recall.} Rule-word retrieval catches about two-thirds of the security rules in a file. In the blind 50-file audit it finds 63 of 95 in-scope reference segments, a recall of 66.3\% (file-clustered bootstrap 95\% confidence interval [48.8, 82.0]). The 32 it misses are rules written as flat statements rather than commands, rules broken across headings or fragments, or rules using words outside the rule-word list. This does not change the denominator of the main estimate; it means the estimate covers rule-word-retrieved security rules, not every security rule in the files.

\noindent\textbf{Error analysis: where the classifier over-credits.} The classifier makes one repeated mistake. It sees a command, file path, or domain that a permission rule could block and counts that as a match, even when the permission rule cannot express the complete condition. The rejected cases are of four kinds. First, a condition the pattern cannot encode, such as ``run npm publish only when cutting a release'' or ``never merge to main, only the user does that,'' where a deny or ask either fires every time or cannot separate the allowed case from the forbidden one. Second, content rather than location, such as ``never commit secrets,'' which no permission pattern inspects. Third, authorization based on meaning or system state, such as ``the backend reads but never writes \texttt{portfolio.locked\_amount},'' which a file-path rule cannot match. Fourth, a data-flow requirement, ``only LLM API calls leave the network,'' which the network allowlist does not express. The strict reference standard rejects all four as partial matches.

\subsection{Coverage by resource family}\label{subsec:family}

\begin{figure}[t]
\centering
\includegraphics[width=\columnwidth]{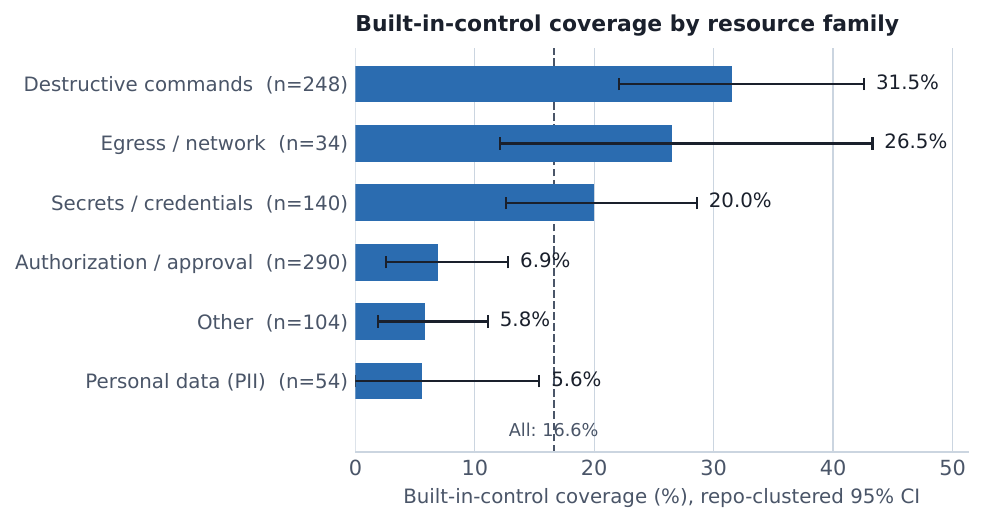}
\caption{Classifier-estimated built-in-control coverage by resource family, with repository-clustered 95\% confidence intervals. These are classifier rates over the full corpus; the exact family ordering is exploratory, with 57.1\% family agreement against the annotator reference.}
\label{fig:family}
\end{figure}

\noindent\textbf{Family-label validation.} The family breakdown is exploratory, because the annotator and the classifier agree only moderately on which family a rule belongs to. On the 91 annotator-reference security-positive rows they pick the same family in 69.2\% of raw cases; correcting for the unequal validation strata, exact agreement is 57.1\% [43.0, 73.2]. Agreement is uneven: all 14 annotator-labeled secrets and all 4 egress rows match, but only 20 of 37 authorization rows do, and the classifier labels 10 of the rest destructive. That one boundary is enough to reorder the families in the small sample, so we read Table~\ref{tab:family} and Figure~\ref{fig:family} as full-corpus classifier descriptions, not as a human-confirmed ranking. The two-annotator agreement (Section~\ref{subsec:reliability}) covers the two binary decisions; we did not run a matched second family pass, so the family comparison stays a single annotator against the classifier.

In the classifier's full-corpus labels, destructive commands and network egress receive the most matches, while authorization and personal data receive the fewest (Table~\ref{tab:family}, Figure~\ref{fig:family}).

\begin{table}[!t]
\centering
\caption{Classifier-estimated built-in-control coverage by resource family, with repository-clustered 95\% confidence intervals.}
\label{tab:family}
\begin{tabular}{lrrl}
\hline
Resource family & $n$ & \% with a match & 95\% CI \\
\hline
Secrets / credentials & 140 & 20.0 & [12.6, 28.6] \\
Egress / network & 34 & 26.5 & [12.1, 43.3] \\
Destructive commands & 248 & 31.5 & [22.1, 42.6] \\
Authorization / approval & 290 & 6.9 & [2.6, 12.8] \\
Personal data (PII) & 54 & 5.6 & [0.0, 15.4] \\
Other & 104 & 5.8 & [1.9, 11.1] \\
\hline
\textbf{All} & 870 & 16.6 & [12.5, 21.4] \\
\hline
\end{tabular}
\end{table}

Table~\ref{tab:examples} gives representative de-identified examples per family. A rule gets a yes only when a named built-in control covers its full action, target, condition, and approval.

\begin{table*}[!t]
\centering
\caption{Representative de-identified examples per family, with the matching-control verdict and the reason.}
\label{tab:examples}
\begin{tabularx}{\textwidth}{@{}l>{\raggedright\arraybackslash}Xc>{\raggedright\arraybackslash}X@{}}
\hline
Family & Example rule (de-identified) & Matching control? & Why \\
\hline
Secrets & ``do not run \texttt{rails credentials}'' & Yes & \texttt{deny} on the Bash command \\
Secrets & ``never store API keys in plaintext JSON in production'' & No & no built-in control reads file content; production condition unobservable \\
Destructive & ``commit and push only when asked'' & Yes & permission \texttt{ask} on the git command \\
Destructive & ``delete the team only after all members have shut down'' & No & the required runtime state is not visible at enforcement \\
Egress & ``programs must execute fully offline once validated'' & No & needs an added network check beyond the domain allowlist \\
Authorization & ``one service module must never use another's database layer'' & No & no built-in control represents module-to-module access \\
Personal data & ``never log passwords, tokens, IBAN, or customer location'' & No & no built-in control inspects log content for these fields \\
Other & ``do not use \texttt{eval()} or assign directly to \texttt{document.cookie}'' & No & a code-content rule; needs a linter or scanner \\
\hline
\end{tabularx}
\end{table*}

\subsection{Coverage by rule wording and by control}

\begin{figure}[t]
\centering
\includegraphics[width=\columnwidth]{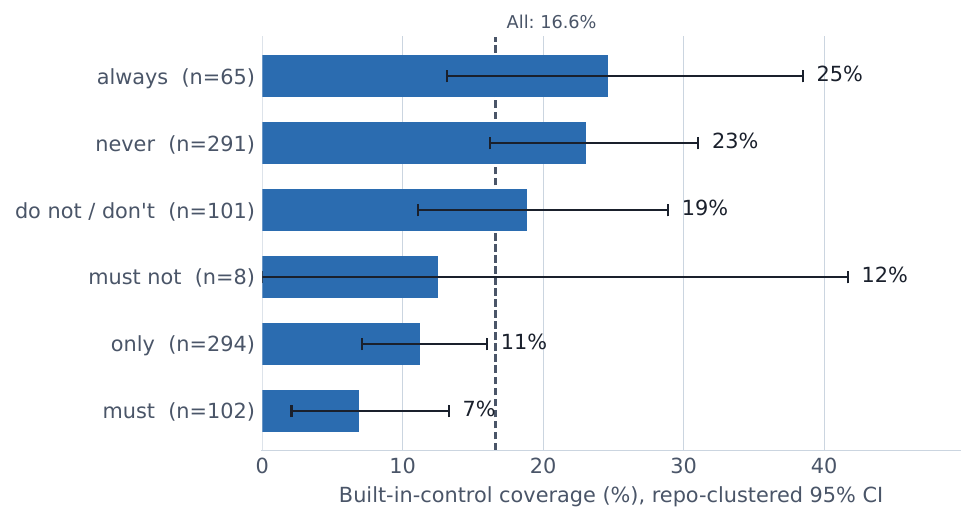}
\caption{Rule wording does not reliably reveal whether a control backs the rule: classifier-estimated coverage by the first rule word is similar and overlapping across the common forms, with repository-clustered 95\% confidence intervals.}
\label{fig:marker}
\end{figure}

\noindent\textbf{Coverage by rule word.} The wording of a rule does not reveal whether it has a matching built-in control. Figure~\ref{fig:marker} breaks the security-rule segments down by the first rule word in the text. Coverage is similar and overlapping across the common forms: \emph{never} ($n{=}291$) at 23\%, \emph{only} ($294$) at 11\%, \emph{must} ($102$) at 7\%, \emph{do not} ($101$) at 19\%, and \emph{always} ($65$) at 25\%, with wide repository-clustered intervals. The figure shows these five and a small \emph{must not} group ($n{=}8$); 9 segments with rarer rule words are not shown. No rule word reliably separates rules with a match from rules without one, which reflects how \texttt{CLAUDE.md} uses the same prose form for both cases.

\begin{figure}[t]
\centering
\includegraphics[width=\columnwidth]{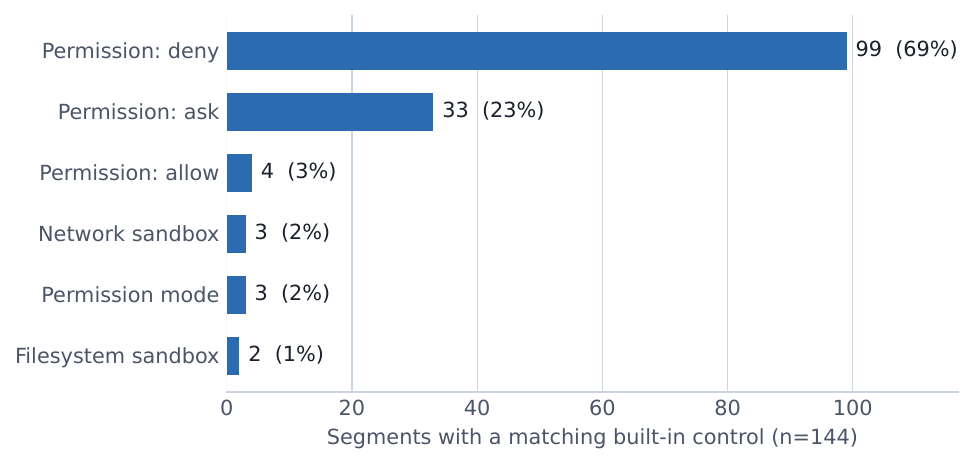}
\caption{Which documented control the classifier cites for the 144 segments it marks as having a match. These controls are not human-verified; the classifier over-credits \texttt{deny} and \texttt{ask}.}
\label{fig:control}
\end{figure}

\noindent\textbf{Which controls the classifier cites.} Among the 144 segments that the classifier marks as having a match, the cited control is almost always a permission rule. Figure~\ref{fig:control} shows 99 (69\%) citing a permission \texttt{deny} rule and 33 (23\%) a permission \texttt{ask} rule; the remaining 12 divide among permission \texttt{allow} (4), the network sandbox (3), permission modes (3), and the filesystem sandbox (2). These are classifier-cited controls, not human-verified matches, and the reliability results in Section~\ref{subsec:reliability} show that the classifier over-credits \texttt{deny} and \texttt{ask} in particular.

For reproducibility, Appendix~\ref{app:split} records the classifier's exploratory reasons for the rules that had no match.

\subsection{Robustness}

\noindent\textbf{Duplicated text and repository concentration.} The low coverage rate is not an artifact of duplicated text or a few large repositories. After normalizing case, formatting, URLs, identifiers, and numbers, only 17 of 870 security segments are additional copies, and the largest template cluster holds 4. Re-weighting to give each template, or each repository, equal weight, and excluding the largest repositories, all leave the estimate in the same range (Table~\ref{tab:sensitivity}). The largest repository contributes 78 segments (9.0\%) and the largest 5 contribute 208 (23.9\%), so concentration matters a little more than duplication, but it does not reverse the result. These are sensitivities of classifier output, not replacements for the validation-adjusted estimate.

\begin{table}[t]
\centering
\caption{Sensitivity of the 16.6\% classifier estimate to duplicate text and repository concentration.}
\label{tab:sensitivity}
\begin{tabular}{lr}
\hline
Weighting or exclusion & Coverage \\
\hline
Rule-weighted (primary) & 16.6\% \\
Each exact template equal & 16.7\% \\
Each near-template cluster equal & 16.6\% \\
Each repository equal & 21.3\% \\
Excluding largest 1 / 3 / 5 / 10 repositories & 17.4 / 17.7 / 19.0 / 18.9\% \\
\hline
\end{tabular}
\end{table}

\begin{figure}[t]
\centering
\includegraphics[width=\columnwidth]{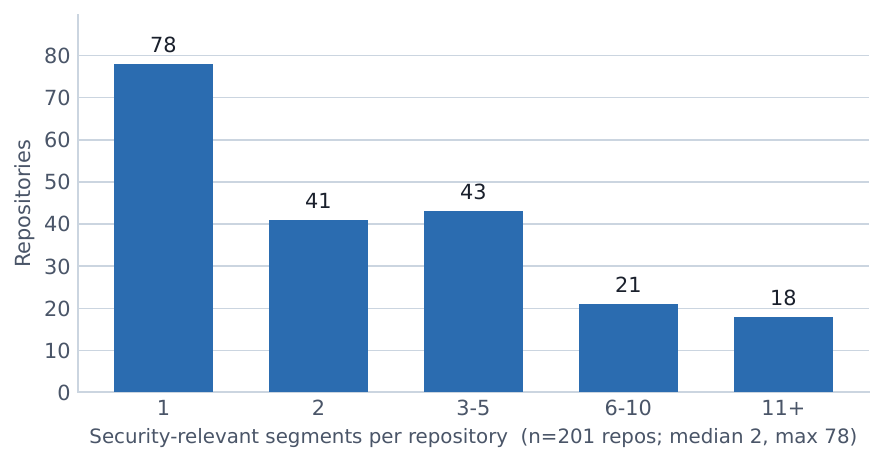}
\caption{Security-relevant segments per repository: a long tail across 201 repositories (median 2, maximum 78). This concentration is why the paper clusters every interval by repository.}
\label{fig:perrepo}
\end{figure}

\noindent\textbf{Corpus structure and scope.} Figure~\ref{fig:perrepo} shows the per-repository distribution: the 870 segments spread across 201 repositories with a long tail (median 2, maximum 78), which is why every interval in this paper is clustered by repository. The corpus skews toward recent, low-visibility projects: 93\% of segments come from repositories created in 2025 and 2026, and 58\% of repositories have no stars, so we do not extend the rate to large, established projects.

\noindent\textbf{A second corpus.} On a second, independent corpus the coverage rate stays a minority, though its exact value depends on the frame. We repeat the full pipeline on the \texttt{CLAUDE.md} subset of the Agent READMEs dataset~\cite{chatlatanagulchai2025}, a published collection of agent instruction files, built by a different method and nearly disjoint from the primary frame (at most three shared repositories). The classifier estimate is higher there than in the primary frame but still a minority (Table~\ref{tab:frames}). The same family pattern replicates, with the gap concentrated in destructive commands (31.5 versus 51.8\%). Because Frame B has no annotator reference labels and Frame A family agreement is modest, this is a sampling-frame replication of instrument output, not independent confirmation of the rate or family ordering.

\begin{table}[t]
\centering
\caption{Classifier coverage estimate in the two sampling frames.}
\label{tab:frames}
\begin{tabular}{lrrr}
\hline
Frame & Segments & Repositories & Coverage (95\% CI) \\
\hline
A (primary) & 870 & 201 & 16.6\% [12.5, 21.4] \\
B (Agent READMEs) & 773 & 701 & 25.1\% [20.8, 30.0] \\
\hline
\end{tabular}
\end{table}

\section{Discussion}
In \texttt{CLAUDE.md}, one written form covers two very different cases: rules a built-in control can apply, and rules only the model can interpret. Our central finding is that only a small minority (about 4 to 16\%, depending on the coding standard) has a matching built-in control, and nothing in the file or the agent application marks which. This makes the file a write-only security interface: a developer states a requirement and gets nothing back. Ordinary software closes this loop with compile errors, failing tests, and runtime observability, so a developer can tell whether an intent took effect; a \texttt{CLAUDE.md} security rule returns no such feedback, so a developer can believe a rule is in force when nothing enforces it. Teams may still apply these rules through mechanisms outside the file, which we do not observe. What the file omits is the security-relevant fact itself: whether a written rule is backed by a control.

The family breakdown is the classifier's view; its agreement with human coding is only moderate (Section~\ref{subsec:family}), so the evidence backs the overall gap, not a ranking among families. A confirmed ranking would need sharper family definitions and a larger sample coded by multiple independent experts.

This gap matters more now that the population writing these rules has changed. Language models have lowered the barrier to building software and agents, and many of the new builders have little security background. Instruction files let them state security constraints in plain prose, without the code review, tests, or continuous integration that used to catch mistakes. Prose is easy to write but hard to verify: there is no type checker for ``never commit secrets.'' Because these agents take real actions, a rule with no control is not cosmetic: the agent may still do what the rule forbids. A write-only interface may be most dangerous for the people least likely to notice that nothing enforces what they wrote.

The design response is to close this loop. An agent platform could show, beside each rule, whether a built-in control backs it; mark rules it cannot enforce as advisory rather than accepting them silently; log when a rule would have applied, so its effect is observable; or help developers turn prose into a reviewable check. Each restores a form of the feedback that ordinary software already provides.

\section{Limitations}
\label{sec:limits}
\noindent\textbf{Scope.} We measure only the rules as written in \texttt{CLAUDE.md}, and only whether Claude Code documents a matching control for them. We do not measure whether a team enforces a rule some other way, through continuous integration, hooks, or organization tooling, whether a repository configured the control, or whether it holds up at runtime. A rule with no matching control is therefore not necessarily unprotected, and our rates do not describe a repository's overall security.

\noindent\textbf{Instrument and annotators.} Our labels come from an LLM classifier, checked against a human reference that two annotators coded independently and then adjudicated, which we do not treat as ground truth. Because the sample draws equally from three unequal classifier strata, every rate uses inverse-probability weights and every interval resamples repositories as clusters. Coding the 180 items twice lets us estimate reliability rather than assert it: before adjudication the two agree on the match decision ($\kappa = 0.89$; Section~\ref{subsec:reliability}) and less on whether a segment is a security rule ($\kappa = 0.77$). The rate still depends most on how strictly a control must match, from about 14 to 17\% under a loose reading to about 4\% under a strict one. We report the 4 to 16\% range and rest the conclusion on what holds across it. Using an LLM as an offline instrument, checked against human labels, is distinct from relying on a model to interpret a rule at runtime.

\noindent\textbf{Family labels.} The family breakdown rests on a single round of human coding with only moderate agreement, so we treat it as exploratory (Section~\ref{subsec:family}).

\noindent\textbf{Generality.} Our results depend on both what developers wrote and the controls Claude Code documented on our freeze date. A post-hoc audit of that control reference found one imprecise entry and one missing credential-protection feature, with no rule that would have matched it, but it cannot prove the reference was complete. The second corpus varies the sampling frame but reuses the same instrument and control vocabulary, so it does not test other products. We therefore bound the product claim to Claude Code and the time claim to the freeze date, and note that public \texttt{CLAUDE.md} files are not a random sample of security-relevant writing.

\noindent\textbf{Classification judgment.} Deciding whether a control matches a rule takes judgment. We constrain it with a frozen prompt, an explicit definition of harm, a requirement that the model name the control it matched, and two blind human checks, and we flag the one boundary that proved contestable: whether a control fully matches a rule or only partly, which is where the two annotators split, on 4 of 180 rules.

\noindent\textbf{Extraction recall.} Our extractor keys on predeclared rule words, so it misses security rules written without them; the 50-file audit puts recall at 66.3\% [48.8, 82.0]. Our rates therefore describe the rules it retrieved, and the audit cannot say whether the missed rules have controls at the same rate.

\section{Conclusion}
Natural language is increasingly how people both build agents and try to secure them. Yet across the different standards we tried, only about 4 to 16\% of the security rules developers write in \texttt{CLAUDE.md} have one, and neither the file nor the agent shows which ones do. A written rule looks like protection, but the channel is write-only: the intent is recorded and never checked. This breaks the feedback loop that has long kept software dependable, where a developer could see whether a requirement took effect. It likely fails hardest for the growing number of builders least equipped to notice. We call on the builders of these tools to treat this as a gap worth closing, and to give natural-language security rules a mechanism that enforces them deterministically where it can, makes that enforcement observable, and warns when a rule is only advice. Closing this loop would make a written rule as dependable as it looks.

\section*{Use of Generative AI}
OpenAI Codex 5.6 Sol assisted with writing code to collect public GitHub repository information, analyze the data, and generate figures in Python. It also assisted with language polishing, verification, and final review. The research questions, study design, coding rules, manual labels, interpretation, and final claims were developed and written by the human authors, who reviewed all outputs and take full responsibility for the paper.

\section{Ethics Considerations}
The study reads only public repository text. It does not run any agent, test any target system, or attempt to bypass any control. We follow the beneficence principle of the Menlo Report~\cite{dittrich2012menlo}: the benefit of measuring how often a written security rule has a matching built-in control outweighs the limited risk of reading public files at rest. We report results in aggregate and do not name individual repositories or authors. We decide whether to publish per-rule source links before release, weighing potential harm, license, and reproducibility. The work involves no participants and collects no personal data.

\bibliographystyle{IEEEtran}
\bibliography{references}

\appendices

\section{Control reference (Claude Code, frozen 2026-08-13)}
\label{app:control}
This table is the shared reference for the classifier and the human coders. A control counts as built-in only when a developer can configure it without writing executable code.

\begin{table*}[!t]
\centering
\caption{Claude Code controls in the frozen reference (2026-08-13) and whether each counts as built-in.}
\label{tab:control}
\begin{tabularx}{\textwidth}{@{}l>{\raggedright\arraybackslash}Xc@{}}
\hline
Control & Action or target it can cover & Built-in? \\
\hline
permission rule: \texttt{deny} & block a tool call, Bash command, Read/Edit path, or WebFetch domain matching a pattern & Yes \\
permission rule: \texttt{ask} & require user approval before a matching action & Yes \\
permission rule: \texttt{allow} & permit a matching action without prompting & Yes \\
permission mode & session-wide approval behavior (default / acceptEdits / plan / bypassPermissions) & Yes \\
\texttt{additionalDirectories} & which directories tools may touch & Yes \\
filesystem sandbox & filesystem reach of Bash subprocesses & Yes \\
network sandbox & network egress of Bash subprocesses (domain allowlist) & Yes \\
managed settings & org-enforced policy users cannot override & Yes \\
\texttt{PreToolUse} / other hooks & author-written executable logic run around tool calls & No (added code) \\
\hline
\end{tabularx}
\end{table*}

This table reproduces the frozen instrument rather than silently repairing it after validation. A 2026-08-16 audit against the official permissions, settings, sandbox, and hooks documentation~\cite{anthropic2026permissions,anthropic2026settings,anthropic2026sandbox,anthropic2026hooks} found that \texttt{additionalDirectories} only grants additional access, that the table omitted \texttt{sandbox.credentials}, and that network denial outside an allowlist requires \texttt{strictAllowlist} or a managed lockdown. \texttt{additionalDirectories} had 0 citations as a matching control. \texttt{sandbox.credentials} had 0 direct text matches among the 870 security-rule segments; manual review of 39 broader credential-scope matches found no rule within its agent-level scope. Permission rules account for 136 of 144 positive citations, permission modes for 3, filesystem sandboxing for 2, and network sandboxing for 3. The released audit CSV records each finding and its observed effect.

\section{Frozen classifier prompt (abridged)}
\label{app:prompt}
One prompt, frozen before validation, applied to every candidate. The released data and frozen prompt use the legacy field name \texttt{first\_class}; in the paper, this means that a matching built-in control exists. We preserve the field name below because it records the prompt actually used. The prompt receives the control reference above, then:
\begin{quote}
\emph{Definitions.} \texttt{security\_relevant} = 1 if violating the line could cause a concrete security or privacy harm (leak a secret/credential/PII, exfiltrate or expose data, unauthorized or irreversible action, weaken authentication/authorization); else 0. \texttt{first\_class} = 1 if a native control above can faithfully bind this rule with no added code (covering its action/resource/condition, not standing in by over-blocking a much larger set); else 0; only meaningful when \texttt{security\_relevant} = 1.

\emph{Rules.} If \texttt{first\_class} = 1, name the exact native control as evidence. Conservative: if unsure whether a native control faithfully covers the rule, set \texttt{first\_class} = 1 (so the reported rate is an upper bound). Judge only this line's text and context; do not assume repo code you cannot see.

\emph{Output.} JSON: \texttt{security\_relevant}, \texttt{first\_class}, \texttt{candidate\_control}, \texttt{resource\_family} (secrets / egress / destructive / authorization / pii / other), \texttt{rationale} (at most 8 words).
\end{quote}
The secondary 4-way outcome is produced by a separate pass (Appendix~\ref{app:split}) over security rules without a matching control, so it cannot alter the binary headline. The prompt's quoted ``upper bound'' phrase records the instrument instruction; the validation results do not establish a one-sided error bound.

\section{Secondary categories for rules without a matching control}
\label{app:split}
The classifier first records whether a matching built-in control exists. For each security rule without a match, it then records one possible reason. The annotator-reference task did not validate these reasons.

\begin{table}[H]
\centering
\caption{Secondary split of the 870 security rules (exploratory).}
\label{tab:split}
\begin{tabularx}{\columnwidth}{@{}>{\raggedright\arraybackslash}Xrr@{}}
\hline
Outcome & $n$ & \% of 870 \\
\hline
matching built-in control & 144 & 16.6 \\
custom (deterministic check writable, needs added code) & 446 & 51.3 \\
model-mediated (needs open-ended judgment) & 181 & 20.8 \\
missing-context (info not observable at enforcement) & 99 & 11.4 \\
\hline
\end{tabularx}
\end{table}

These counts are retained only to reproduce the frozen classifier output and generate future hypotheses. Because no annotator validated these reasons, we treat their percentages as exploratory and do not read the ``custom'' share as the fraction of the gap that added code could close.

\section{Reproducibility}
\label{app:repro}
\begin{itemize}
\item \textbf{Primary frame}: GitHub code search for filename \texttt{CLAUDE.md}, retrieved 2026-08-14; 647 repositories examined, 481 included after the inclusion filters (Section~\ref{sec:split}).
\item \textbf{Replication frame}: the \texttt{CLAUDE.md} subset of the Agent READMEs dataset~\cite{chatlatanagulchai2025}; 701 included.
\item \textbf{Seed}: 20260813 (sampling and all bootstrap resampling). \textbf{Classifier}: Claude Sonnet 5, frozen prompt. \textbf{Validation}: 180 candidates labeled blind to classifier outputs, independently by two security practitioners (the second added post-hoc; each blind to the classifier and to the other), with disagreements adjudicated against the frozen control reference to form the consensus reference; the 91 security-positive rows also receive a single blind resource-family pass. \textbf{Extraction-recall audit}: 50 complete files, 95 in-scope reference segments after scope re-adjudication, with file-clustered bootstrap intervals.
\item \textbf{Released}: the frozen sampling frames, per-frame manifests (repository, path, commit SHA), the control reference and post-hoc audit, the extraction/classification/analysis scripts (including duplicate/template and repository-concentration sensitivity), aggregate outputs, and \texttt{references.bib}. Per-rule source links are reviewed for harm and license before release (Section~\ref{sec:limits}).
\end{itemize}

\end{document}